# Direct Observation of an Interface Dipole between Two Metallic Oxides Caused by Localized Oxygen Vacancies

A.Y. Borisevich,[1,*] A.R. Lupini,[1] J. He,[1,2] E.A. Eliseev,[3] A.N. Morozovska,[4],

G.S. Svechnikov,[4] P. Yu,[5] Y.H. Chu,[6] R. Ramesh,[5] S. T. Pantelides,[2,1] S.V. Kalinin,[1] and S.J. Pennycook, [1,2]

[1] Oak Ridge National Laboratory, Oak Ridge, TN 37831

[2]Department of Physics and Astronomy, Vanderbilt University, Nashville, TN

[3] Institute for Problems of Materials Science, National Academy of Science of Ukraine, 3, Krjijanovskogo, 03142 Kiev, Ukraine

[4] Institute of Semiconductor Physics, National Academy of Science of Ukraine, 41, pr. Nauki, 03028 Kiev, Ukraine

[5]Department of Materials Science and Engineering and Department of Physics, University of California, Berkeley, California, 94720

[6] Department of Materials Science and Engineering, National Chiao Tung University, Hsinchu, Taiwan 30013 (ROC)

---

[*] Corresponding author, albinab@ornl.gov

Oxygen vacancies are increasingly recognized to play a role in phenomena observed at transition-metal oxide interfaces.  Here we report a study of $SrRuO_3$/$La_{0.7}Sr_{0.3}MnO_3$ (SRO/LSMO) interfaces using a combination of quantitative aberration-corrected scanning transmission electron microscopy, electron energy loss spectroscopy, and density-functional calculations. Cation displacements are observed at the interface, indicative of a dipole-like electric field even though both materials are nominally metallic. The observed displacements are reproduced by theory if O vacancies are present in the near-interface LSMO layers. The results suggest that atomic-scale structural mapping can serve as a quantitative indicator of the presence of O vacancies at interfaces.

Emergent electronic and structural phenomena at transition-metal oxide (TMO) interfaces have become one of the key areas of condensed matter physics. This interest is driven strongly by applications necessitating development of materials with novel superconducting,[1] transport,[2] and magnetoelectric functionalities.[3] These applications are enabled by a broad spectrum of physical and chemical phenomena due to charge mismatch,[2] polarization discontinuities,[4] orbital ordering,[5] spin reconstruction,[6,7,8] ionic transfer and vacancy segregation,[9,10] or octahedral tilt effects.[11,12] One of the key aspects of oxide interfaces is the multiple electronic, magnetic, structural, and chemical mechanisms that can be operational at the same time. This variety of effects can greatly complicate the identification of the individual physical phenomena,[13] and necessitates comprehensive studies of all aspects of interfacial behavior.[14]

The role of oxygen vacancies in the above phenomena has been largely unexplored. It is well known that many materials such as manganites, cobaltites, and nickelates have large (3-10%) concentrations of oxygen vacancies that strongly depend on oxygen activity. Mitchell et al. have demonstrated that relatively small changes in oxygen partial pressure (4 orders of magnitude in $pO_2$ corresponding to ~25-50 meV in electrochemical potential) can affect the phase diagram of LSMO similar to 25% doping.[15] These effects can be expected to be even more pronounced at interfaces with large build-in electric fields. Recent studies by the groups of Skowronski[16] and Cheong[17] have demonstrated that vacancies on the nanoscale can be mobile even at room temperature, and hence cannot be assumed to be frozen. This recognition of the role of vacancies is contrasted with the lack of local observations; indeed, only in the cases of well-established vacancy ordering, such as mixed-valence cobaltites[18] and high-temperature superconductors[19] can the oxygen vacancies be unequivocally detected.

Density functional calculations, however, in conjunction with experimental data can offer strong evidence for their presence or absence.[9]

In this Letter, we explore the structural and electronic behavior at the interface between the $SrRuO_3$ (SRO) and $(La_{0.7}Sr_{0.3})MnO_3$ (LSMO) using quantitative aberration corrected scanning transmission electron microscopy (STEM) and electron energy loss spectroscopy, extending the approach suggested for transmission electron microscopy (TEM) by Jia[20,21] that has been recently adapted for STEM.[9,12,22,23] The cation displacement profile at the nominally metal-metal SRO-LSMO interface is visualized directly. The observed displacements can be interpreted as indicative of an interfacial electric dipole We show that the observed cation displacement pattern can be reproduced quantitatively by density functional theory if oxygen vacancies are present at the interface. Furthermore, the electronic potential profiles obtained from DFT and extracted from the STEM data by phenomenological modeling match closely. The implications for the electronic and magnetic properties are discussed.

LSMO thin films (with the thickness of 6nm) were grown on (001) single crystal STO substrate using pulsed laser deposition at 700°C and an $O_2$ pressure of 300 mTorr. Subsequently, the SRO layer (with the thickness of 30 nm) was grown on LSMO buffered STO at 700 °C and with an $O_2$ pressure of 100 mTorr. HAADF STEM imaging and EELS studies were carried out using VG Microscopes HB603U operated at 300 kV and equipped with a Nion® aberration corrector and Gatan Enfina® spectrometer.

The HAADF STEM image of the SRO-LSMO interface is shown in Fig. 1 (a), demonstrating a clear transition between LSMO and SRO. In this mode, atomic column intensity is roughly proportional to the square of atomic number Z, so constituent elements can be deduced by tracking column intensities.[24] Analysis of A-site and B-site intensities across the SRO/LSMO interface for multiple images shows that at the transition columns with intermediate intensities are often present, on the scale of up to one unit cell for Mn/Ru sublattice and on the scale of 0-2 unit cells for Sr/La sublattice, similar to that observed by Ziese *et al.*[25] in SRO/LSMO multilayers. Along the interface the films appear uniform, with no evidence of the extended defects reported to arise in non-optimal growth conditions.[26] The number of intermediate intensity columns is reduced for thinner areas of the sample, suggesting that SRO surface steps (in the beam direction), rather than uniform intermixing, are the likely cause of the intermediate column intensities. With that in mind, we have chosen a thin sample area for our quantitative examination where the results would not be affected by vertical averaging of different atom types. The interface was also analyzed by Electron Energy Loss Spectroscopy (EELS). The profiles of La and Mn across the interface show behavior consistent with HAADF intensity traces (Fig.1(b)). Note that (a) the La signal extends beyond the Mn signal at the interface, indicating effective (La,Sr)O/$RuO_2$ termination and (b) the widths of the profiles are identical, indicating no or equivalent intermixing between both cationic sublattices.

The details of the interface structure were investigated by direct atomic position mapping.[9,12] Typical maps of *c*-lattice parameter (normal to the interface) are shown in Fig. 1 (c). The *c*-parameter map shows that saturation lattice parameter decreases in LSMO, in agreement with the difference in bulk lattice parameters. It also shows some anomalies at the interface, as can be more clearly seen from the averaged profile in Fig. 1 (d). The ~ 1 u.c. thick layer of SRO is slightly compressed approaching the interface, whereas the LSMO component shows expansion over ~ 3 u.c.

The corresponding map and profile of a-lattice parameters (parallel to the interface) are shown in Fig. 1 (e, f). While the average a parameter is constrained to the same value on both sides of the interface by epitaxy, the lattice parameter map in the SRO shows clear checkerboard pattern, indicative of the orthorhombic, rather than cubic phase. This observation is in agreement with recent studies by the Eom group.[27] This orthorhombic distortion is however not expected to significantly affect the electronic and magnetic structure of SRO.[28] Note that the checkerboard pattern is non-uniform along the interface, and nominally cubic regions can penetrate by 3-4 unit cells in the SRO layer, indicating a complex strain-related behavior.

To complement lattice spacing maps, we show a cation displacement map, in which displacements are calculated as the distance between the midpoint of the two adjacent Sr/La sublattice columns and the corresponding Ru/Mn column. The X (perpendicular to the interface) and Y (parallel to the interface) components of displacement are shown in Fig. 2 (a,b); the corresponding averaged profiles are shown in Fig. 2 (c). In-plane displacement (Y) is near-zero and shows a small change on transition to LSMO, possibly due to a small relative mis-rotation of the LSMO and SRO blocks. At the same time, the X displacement map shows an anomaly at the interface, with negative displacement in the SRO region extending ~3 u.c. inside the material, and positive displacement in the LSMO region extending ~6. u.c. into the material; in our sign convention, both displacement constitute motion towards the interface. This behavior is highly unusual, given that both materials are nominally in the metallic state (SRO is a majority-electron metal; LSMO is a majority-hole semi-metal), and suggests the presence of built-in electric field at the SRO-LSMO junction.

To estimate the potential distribution in the vicinity of the interface, we adopt a phenomenological Ansatz and assume that the displaced B cations (Mn and Ru) carry a nominal positive charge $q_\alpha$, where $\alpha=1, 2$ for the two sides of the interface. As in Refs. 20, 23, we write the resulting polarization field on the two sides as

$$P_\alpha(z) = q_\alpha u_\alpha(z) / V_\alpha \qquad (1.1)$$

where $V_\alpha$ are unit cell volumes, $u_\alpha(z)$ are the relative displacements of the B cations (see Fig. 2), and $z$ is the distance across the interface (with the interface at $z=0$). To determine the interfacial fields and potentials, we treat $P_\alpha(z)$ as a continuous function, which is in effect an interpolation of the original discrete dataset. To avoid artifacts from numerical differentiation of noisy input, we model $P_\alpha(z)$ by assuming that it arises from an effective charge density $\rho_\alpha^f(z)$ which we express as a sum of a finite number of step functions,

$$\rho_\alpha^f(z) = \sum_i \rho_{\alpha i} \theta(|z| - z_{\alpha i}) \theta(z_{\alpha i} + W_{\alpha i} - |z|)$$

. The electrostatic potential $\varphi_\alpha(z)$ on each side of the interface is given by $\varepsilon_0 \varepsilon_\alpha \partial^2 \varphi_\alpha(z) / \partial z^2 = -\rho_\alpha^f(z)$, where the $\varepsilon_\alpha$ are (bare) lattice permittivities. While SRO and LSMO are metal and semimetal, respectively, the two materials can be treated as semiconductors for the purpose of this derivation.[29] The values of the lattice permittivities were taken from taken from refs. 30, 31 for SRO and extrapolated for LSMO from the values for LaMnO$_3$ and SrMnO$_3$.[32] We enforce interfacial boundary conditions $\varphi(+0) = \varphi(-0) = U_b$ and $\varepsilon_0 \varepsilon_1 E(-0) - \varepsilon_0 \varepsilon_2 E(+0) = 0$, where $E(z) = -\partial \varphi(z)/\partial z$ and $U_b$ is the difference in electron affinities in the two materials. The polarization field $P_\alpha(z)$ on the two sides is then given by $P_\alpha(z) = \varepsilon_0(\varepsilon_\alpha - 1)E_\alpha(z)$. In this way, $P_\alpha(z)$ is a function of the $\rho_{\alpha i}$. We then treat the $\rho_{\alpha i}$, $W_{\alpha i}$, and $z_{\alpha i}$ as free parameters and determine the best fit of the function $P_\alpha(z)$ to the discrete values given by the experimental data through Eq. (1.1). This procedure gives us the optimized functions $P_\alpha(z)$, $E_\alpha(z)$, and $\varphi_\alpha(z)$. To quantify the results, we adopted Bader charges for the $q_\alpha$ (calculated as 1.67 and 1.62 for Ru and Mn, respectively, as in Ref. [33]). The results, with an overlay of recalculated experimental data, are plotted in Fig. 3(a-c). Since the profiles showed very weak dependence on the interface potential difference $U_b$, it was set to zero. We note that the electrostatic potential in Fig. 3(c) has a pronounced dipolar character.

While the potential profile in Fig. 3(c) is quite wide, the related cation displacements are localized at the interface. Broadly speaking, the displacement of both Ru ions in SRO and Mn ions in LSMO towards the interface implies electron doping by either free electrons or localized oxygen vacancies. To elucidate the atomic configuration most consistent with the observed cation displacements and dipolar character of the interface, we conducted a density functional theory (DFT) study. Calculations were performed within the framework of the plane-wave basis set and projector-augmented-wave method implemented in the Vienna *ab initio* Simulation Package (VASP)[34,35] The spin-polarized generalized-gradient approximation is applied to the ground state structures of LSMO (rhombohedral, space group $R\bar{3}c$) and SRO (orthorhombic, space group *Pnma*). The Sr doping is considered by the substitution of Sr at La sites with a ratio of 1:3 and the magnetic states are relaxed in collinear configurations. Through structure relaxation, the magnetic coupling at LSMO/SRO interfaces on a SrTiO$_3$ substrate is found to be antiferromagnetic, which is in agreement with a previous DFT study.[25]

To consider intrinsic free-electron doping at the interface (an effect often considered for oxide heterostructures such as LAO/STO[36]), we first constructed interface models by joining stoichiometric LSMO and SRO with the experimentally observed $RuO_2$/La(Sr)O termination. In this case, the La ions near the interface donate electrons to the interface and the interface is therefore doped by free electrons. However, the calculated B-site displacements with $RuO_2$/La(Sr)O termination are relatively small compared to the experimental data (Fig. 4(a), black squares), which is to be expected given that both LSMO and SRO are nominally metallic (unlike STO or LAO), and thus capable of effectively screening interface charge with mobile charge carriers. Thus, the observed B-site polar displacements cannot be ascribed to chemical bonding effects and cation mismatch at the interface.

Based on the fact that no changes in composition and connectivity of the cation sublattice are visible by STEM, this leaves oxygen vacancy segregation as a possible source of the observed phenomena. To explore this possibility, DFT calculations were performed for multiple vacancy distribution scenarios. The results show that introducing even a single oxygen vacancy into the LSMO model near the interface leads to significant cation displacement (Fig.4(a), red squares); the shape of the profile becomes similar to the experimental data (see Fig. 2(c)), indicating qualitative agreement. When two oxygen vacancies are introduced into the near-interface region of LSMO (Fig. 4(a), blue squares), the overall shape of the profile is preserved, while the difference between the two extreme displacements (~0.11 Å) becomes comparable with the experimentally observed value of 0.12 Å (this concentration of vacancies corresponds to one quarter of a monolayer areal coverage of the interface plane). Note that when we introduce oxygen vacancies into the near-interface region of SRO, the resulting cation displacements are much smaller (not shown). Finally, we also considered different possible vacancy sites in LSMO. When the vacancies are moved away from the interface, the displacements become much smaller, suggesting that the experimentally observed B-site polar displacement likely originates from the electron redistribution following introduction of multiple strongly localized oxygen vacancies in the LSMO near the LSMO/SRO interfaces. Thus, the structure model shown in Fig. 4(b) provides the best agreement with the experimental data.

Using these results, we evaluate the potential jump across the LSMO-SRO interface by calculating the electrostatic potentials experienced by the electrons in both the perfect LSMO-SRO supercell model and the model with two oxygen vacancies. To generate profiles across the interface, we performed macroscopic planar averaging of the DFT electrostatic potential along the z direction, followed by a sliding-window averaging of the difference between these two planar-averaged electrostatic potentials (no vacancies and two vacancies). The averaging window used was double the average lattice constants of these oxides. The resulting profile is given in Fig. 4(c); note the characteristic dipolar shape. Notably, to compare this figure to Fig. 3(c), one needs to multiply it by -1, as Fig.4(c) gives a potential experienced by an electron, while Fig. 3(c) gives a potential experienced by e unit positive charge.  Note that both the magnitude and spatial localization of electric fields reconstructed from STEM data in Fig. 3(c) and from DFT are quite close; the full numerical equivalence cannot be expected given that the estimate in Fig. 3(c) did not take the contribution of oxygen anion displacements into account (however, as the DFT calculations show, the O displacements are largely rotations). We also note that the finding that oxygen vacancies are causing the observed displacements does not invalidate the phenomenological model that we employed to describe the macroscopic polarization and potential at the interfaces, since the vacancies cause the effective electric field that displaces the cations. Thus, we can conclude that direct image analysis of the STEM data allows us to obtain polarization and electric fields in the material.

We further note that a large concentration of oxygen vacancies could have produced a measurable change in EELS spectra, either in the integrated signal of the O K edge, or in its fine structure. However, on transition from LSMO to SRO both the integrated O K signal and the "pre-peak" intensity undergo significant changes (~50% total), thus effectively masking any interface-specific contributions. At the same time, much more sensitive atomic displacement mapping allows us to uncover the underlying behavior quantitatively.

To summarize, we explored the atomic structure of the SRO-LSMO interface using direct structural imaging by STEM. This approach allows us to analyze the behavior at the SRO-LSMO interface, complementing recent work Ref. [25], in which magnetic coupling was studied and ascribed predominantly to Mn/Ru intermixing. We attribute the observed phenomena to oxygen vacancy segregation at the interface, as suggested by the excellent agreement between the STEM data and DFT modeling. We further obtain quantitative agreement between electrostatic fields reconstructed from STEM data and from DFT. Beyond the LSMO-SRO system studied here, we note that this structural analysis can be extended to other oxide interfaces, complementing the traditional EELS imaging studies and providing secondary structural parameters of charge transfer, ionic effects, and order parameter couplings across correlated oxide interfaces.

The research is sponsored by the Division of Materials Sciences and Engineering, Office of Basic Energy Sciences, U.S. Department of Energy (AYB, JH, STP, ARL, SJP). Research at Berkeley was sponsored by the SRC NRI- WIN program. Y.H.C. acknowledges the support of the National Science Council, R.O.C., under contract No. NSC 100-2119-M-009-003.

**Figure captions**

**Figure 1.** (a) HAADF image of the SRO-LSMO interface (b) integrated intensities of La M4,5 and Mn L2,3 edges across the interface illustrating interface sharpness and effective (La,Sr)O-RuO$_2$ termination. (a,b) 2D maps of the (a) out-of-plane, or *c* and (b) in-plane, or *a* pseudocubic lattice parameters computed from Fig.1(a). (c,d) Profiles of the maps in (a,b) calculated by averaging along the interface.

**Figure 2.** (a,b) 2D maps of the (a) out-of-plane, or X and (b) in-plane, or Y Mn/Ru cation displacements computed from Fig.1(a).(c) Profiles of the maps in (a,b) calculated by averaging along the interface.

**Figure 3.** Polarization (a), electric field (b), potential (c) and effective charge density (d) profiles reconstructed from experimental atomic displacement data. Filled symbols in (a-d) are calculated from experimental atomic displacement data (Fig. 2(c)). Solid curves are calculated self-consistently for material parameters in text.

**Figure 4.** (a) Mn/Ru cation displacement profiles generated from density functional calculations for (La,Sr)O-RuO$_2$ terminated surface: free-electron doped ( black squares), with one oxygen vacancy on the LSMO side ( red squares), and with two oxygen vacancies on the LSMO side (blue squares). (b) Structure model with two oxygen vacancies (denoted by red circles) showing the best agreement with the experiment, and the difference map of electron

density (for models with and without vacencies); (c) The difference between planar averaged electric potentials for the case with two vacancies (model in Fig.4(b)) vs. no vacancies.

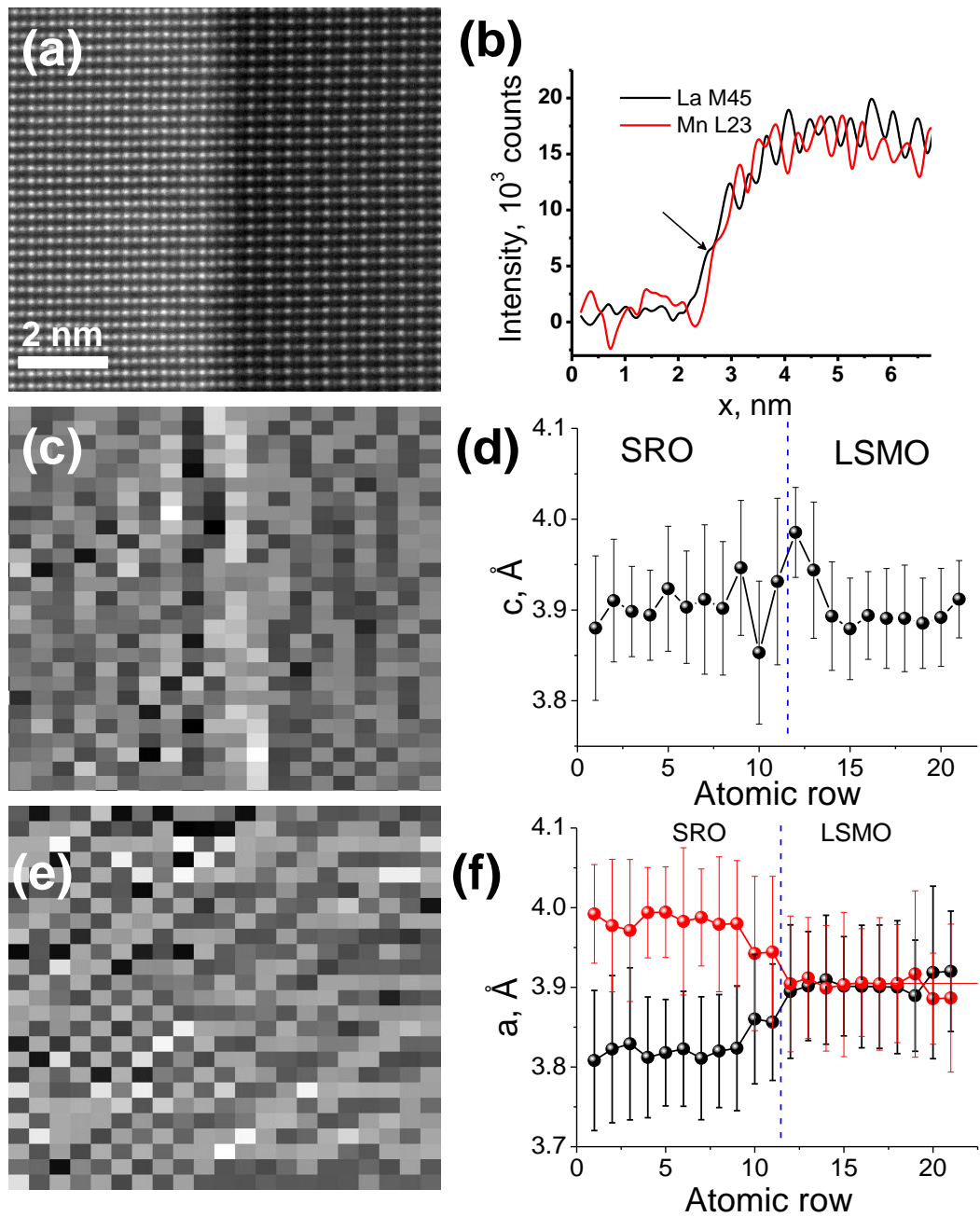

Fig. 1

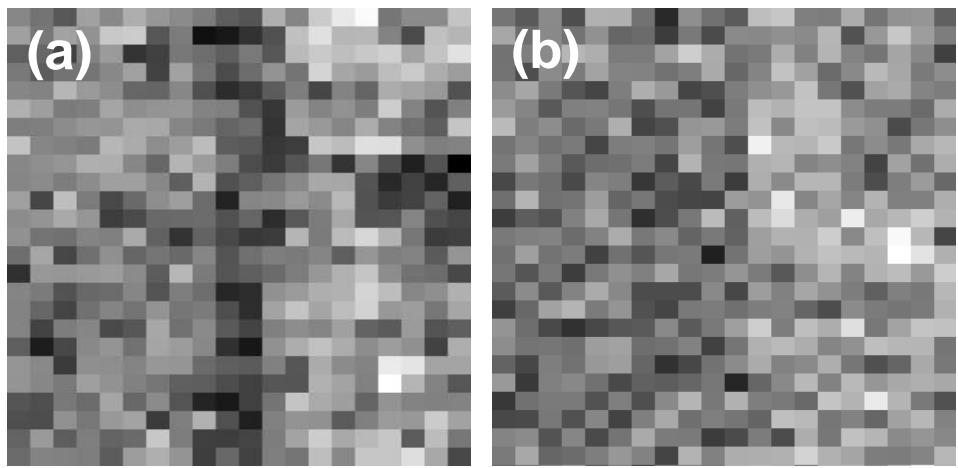
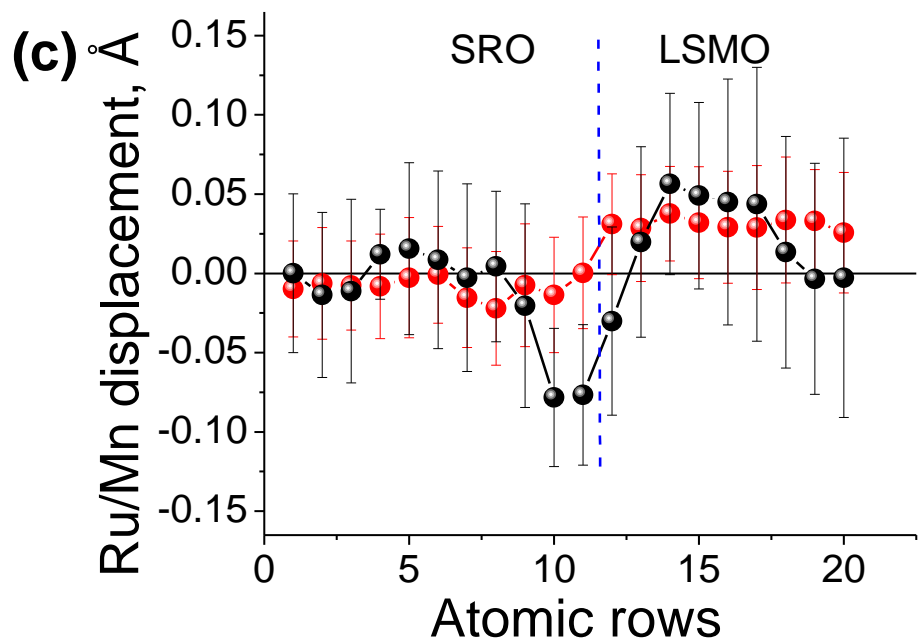

Fig.2

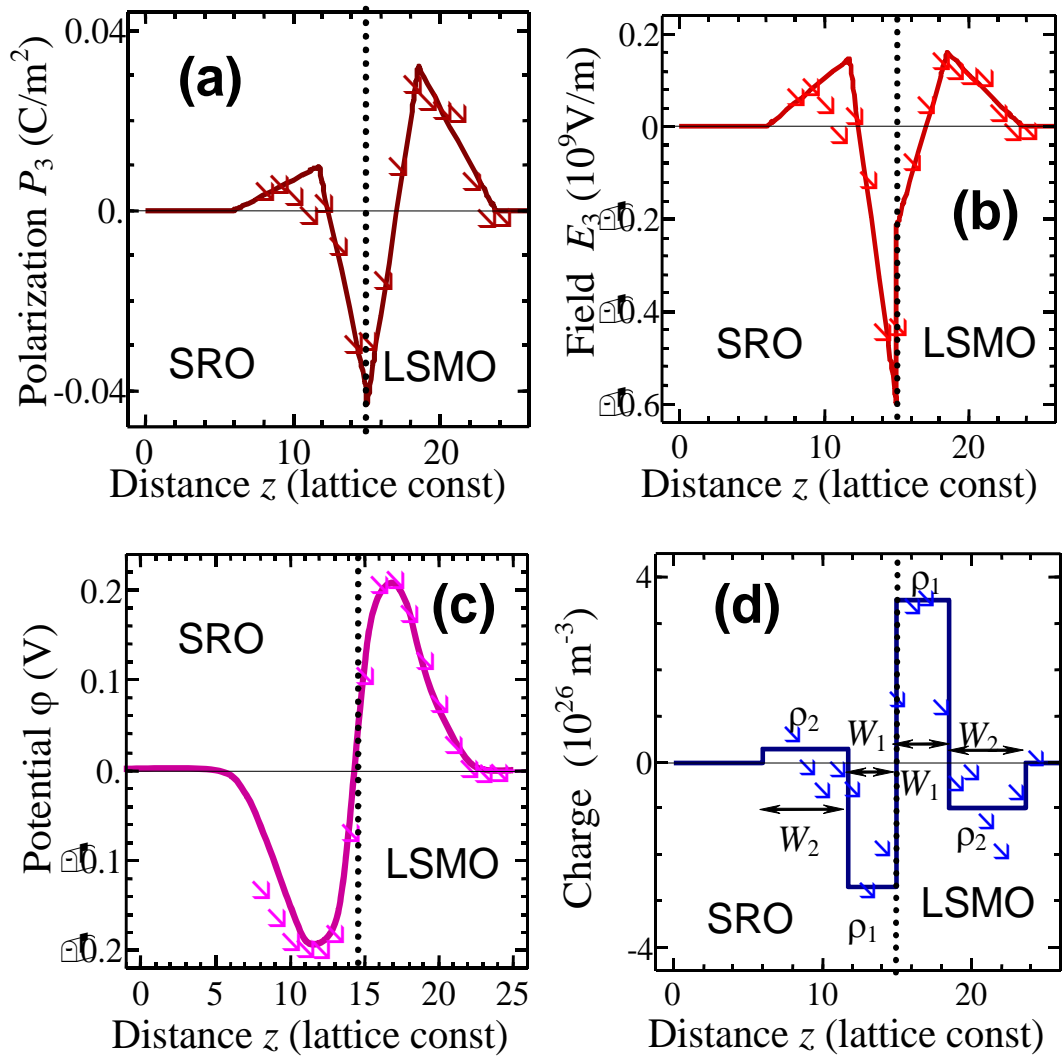

Fig.3

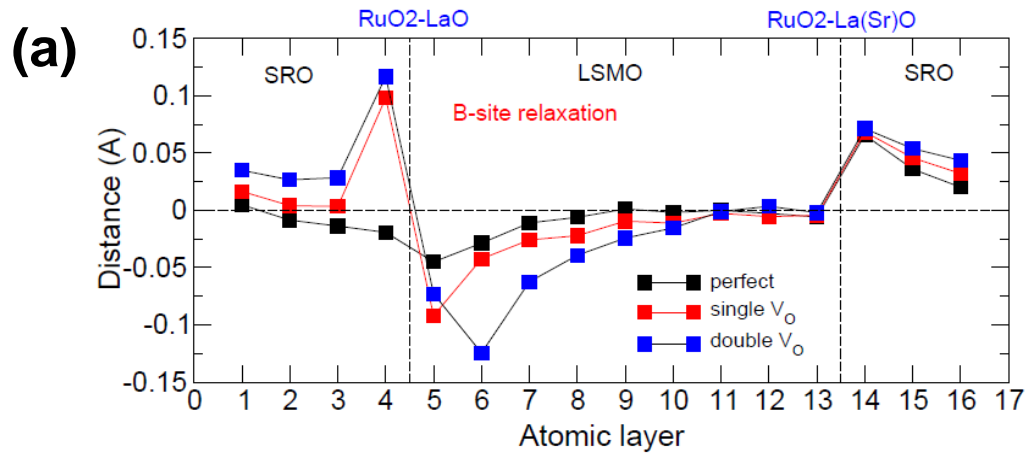
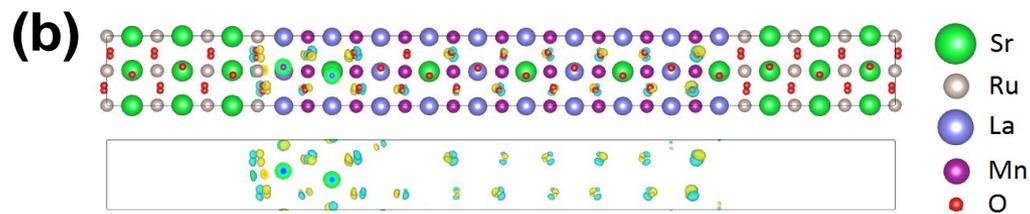
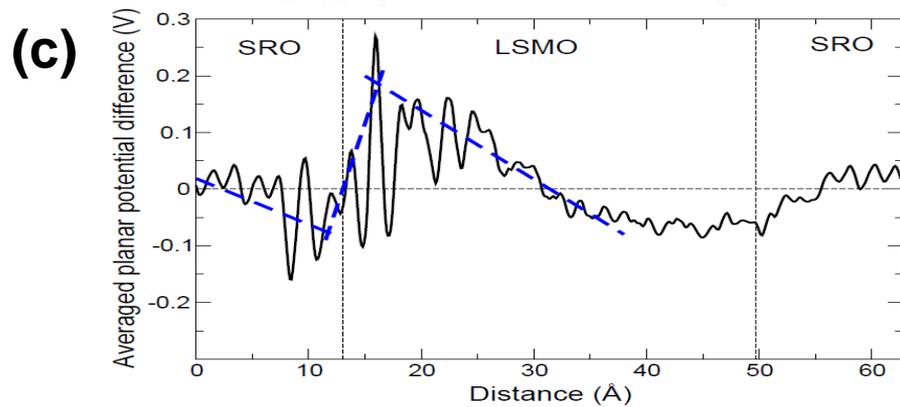

Fig.4